\documentclass{article}
\usepackage[latin9]{inputenc}
\usepackage{geometry}
\usepackage{amsmath}
\usepackage{amssymb}
\usepackage[unicode=true,pdfusetitle,
 bookmarks=true,bookmarksnumbered=false,bookmarksopen=false,
 breaklinks=false,pdfborder={0 0 1},backref=section,colorlinks=false]
 {hyperref}

\makeatletter

\providecommand{\tabularnewline}{\\}

\usepackage{float}
\usepackage{graphicx}
\usepackage{grffile}
\usepackage{breakurl}

\providecommand{\tabularnewline}{\\}

\usepackage{fullpage}\usepackage[justification=centering]{caption}\usepackage[all]{hypcap}\usepackage{microtype}

\makeatother

\begin{document}

\title{Degeneracies with the Q.Q interaction in a single j shell }

\author{Larry Zamick and Alberto Escuderos\\
 \textit{Department of Physics and Astronomy, Rutgers University,
Piscataway, New Jersey 08854}\\
 }

\date{\today}

\maketitle

\section{Abstract}

Previously {[}1{]} it was shown that for a configuration of 2 protons
and 2 neutrons in the g$_{9/2}$ shell there is a certain degeneracy
that occurs when the quadrupole-quadrupole (Q.Q)interaction is used
to obtain wave functions. We here show 3 other examples of such degenerate
T=0,T=2 pairs. More importantly we note a peculiarity of the T=0 partner
in the original example (g$_{9/2}$,J=4 T=0). Also we point out that
degeneracies can be confusing and steps can be taken to remove them.

\section{Introduction }

In a work for a proceedings honoring Aplodor Raduta {[}1{]} Escuderos
and Zamick discussed several topics.One of these involved surprising
degeneracies that occur in single j shell calculatiotns in which he
quadrupole quadrupole (Q.Q) interaction is used to obtain the wave
functions .of a system of 2 protons and 2 neutrons in a single j shell.
It was found that two J=4$^{+}$ states were degenerate- one with
isospin T=0 and the other with isospin T=2. There was no obvious explanation
for this . In this work we do not offer and explanation either but
we systematically search for other examples. As will soon be shown
we find three others also involving one state with T=0 and the other
with T=2. This suggests that this behavior is no accident. We will
find that in one seniority comes into play in determining the properties
of the degenerate pair.

\section{List of degenerate states and their wave functions.}

In tables I,II,II and IV we show the energies and wave functions of
degenerate pairs (T=0 and T=2) for a sytem of 2 protons and 2 neutrons
in a single j shell. The wave functions are represented by column
vectors with entries D$^{J\alpha}$(J$_{P}$,J$_{N}$), the latter
being the probability amplitude that in the $\alpha$'th state of
total angular momentum J the protons couple to J$_{P}$and the neutrons
to J$_{n}$. Note that J$_{P}$and J$_{N}$ are both even. For brevity
we only list the amplitudes where J$_{P}$$\leqq$J$_{N}$. We can
infer the others from the relation

D$^{J,T}$(J$_{P},$J$_{N}$)= (-1)$^{s}$ D( J$_{N}$J$_{P}$) with
s= J+T.

.

.Table I f$_{7/2}$shell, J=2$^{+}$ E= 3.604 MeV.

\begin{tabular}{|c|c|c|}
\hline 
J$_{P,}$J$_{N}$ & T=0 & T=2\tabularnewline
\hline 
\hline 
0, 2 & 0.0832 & -0.2877\tabularnewline
\hline 
2,2 & -0.2689 & -0.2694\tabularnewline
\hline 
2,4 & -0.3510 & 0.3869\tabularnewline
\hline 
4,4 & 0.1212 & -0.0431\tabularnewline
\hline 
4,6 & -0.2115 & 0.3138\tabularnewline
\hline 
6,6 & 0.7507 & 0.5125\tabularnewline
\hline 
\end{tabular}

.

.Table II g$_{9/2}$shell J=4$^{+}$ E=3.5284 MeV.

\begin{tabular}{|c|c|c|}
\hline 
J$_{P,}$,J$_{N}$ & T=0 & T=2\tabularnewline
\hline 
\hline 
0,4 & 0 & 0\tabularnewline
\hline 
2,2 & 0.3132 & -0.4270\tabularnewline
\hline 
2,4 & 0.2289 & -0.2542\tabularnewline
\hline 
2,6 & -0.2076 & 0.3107\tabularnewline
\hline 
4,4 & -0.0135 & 0.2395\tabularnewline
\hline 
4,6 & 0.1784 & -0.1418\tabularnewline
\hline 
4,8 & -0.0888 & 0.1567\tabularnewline
\hline 
6,6 & 0.1362 & 0.1638\tabularnewline
\hline 
6,8 & 0.1353 & 0.0316\tabularnewline
\hline 
8,8 & 0.7549 & 0.5665\tabularnewline
\hline 
\end{tabular}

.

.Table III g$_{9/2}$ shell J=5$^{+}$ E= 4.5890 MeV.

\begin{tabular}{|c|c|c|}
\hline 
J$_{P,}$J$_{N}$ & T=0 & T=2\tabularnewline
\hline 
\hline 
2,4 & 0.0542 & 0.3711\tabularnewline
\hline 
2,6 & 0.2983 & 0.1707\tabularnewline
\hline 
4,4 & 0 & 0\tabularnewline
\hline 
4,6 & 0.3133 & -0.2679\tabularnewline
\hline 
4,8 & -0.4809 & -0.2371\tabularnewline
\hline 
6,6 & 0 & 0\tabularnewline
\hline 
6,8 & -0.2805 & -0.4531\tabularnewline
\hline 
8,8 & 0 & 0\tabularnewline
\hline 
\end{tabular}

.

.Table IV h$_{11/2}$ shell J=5$^{+}$E= 4.1458 MeV

\begin{tabular}{|c|c|c|}
\hline 
J$_{P}$, J$_{N}$ & T=0 & T=2\tabularnewline
\hline 
\hline 
2,4 & 0.0649 & 0.3818\tabularnewline
\hline 
2,6 & -0.1946 & 0.0649\tabularnewline
\hline 
4,4 & 0 & 0\tabularnewline
\hline 
4,6 & -0.3041 & -0.1749\tabularnewline
\hline 
4,8 & 0.1981 & 0.2784\tabularnewline
\hline 
6,6 & 0 & 0\tabularnewline
\hline 
6,8 & -0.0217 & -0.2382\tabularnewline
\hline 
6,10 & 0.3756 & 0.1319\tabularnewline
\hline 
8,8 & 0 & 0\tabularnewline
\hline 
8,10 & 0.4309 & -0.4096\tabularnewline
\hline 
10,10 & 0 & 0\tabularnewline
\hline 
\end{tabular}

. In the above the Q.Q interaction was scaled so that the J=0 two-body
matrix element was set equal to -1.0 MeV.

We now comment on the zeros that appear in the tables. The ones for
odd J i.e. J=5 are easy to understand. They follow from Eq 1 above.
We have

D$^{5}$(J$_{P}$J$_{N}$)= - D$^{5}$(J$_{N}$,J$_{P}$) (-1)$^{T}$.Hence
we have the result, D$^{5,0}$(J,J)=0.

We will discuss the zeros for g$_{9/2}$ J=4$^{+}$ in the next section.

TableV Unique J=4$^{+}$v=2 wave function.

.%
\begin{tabular}{|c|c|}
\hline 
J$_{P}$ J$_{N}$ & D(J$_{P},J$$_{N}$)\tabularnewline
\hline 
\hline 
0,4 & 0.6325\tabularnewline
\hline 
2,2 & 0.1130\tabularnewline
\hline 
2,4 & -.0518\tabularnewline
\hline 
2,6 & -.1129\tabularnewline
\hline 
4,4 & -.0419\tabularnewline
\hline 
4,6 & 0.0970\tabularnewline
\hline 
4,8 & 0..0680\tabularnewline
\hline 
6,6 & -.1725\tabularnewline
\hline 
6,8 & 0.2189\tabularnewline
\hline 
8,8 & -.0311\tabularnewline
\hline 
\end{tabular}

\section{Partial dynamical symmetery-a new feature}

The zeros for g$_{9/2}$ J=4 were recognized in ref{[}1{]} to be connected
to a partial dynamical symmetry. If we focus on T=2 states, we note
that they are double analogs of states of 4 idential particles--say
4 neutrons. For J=4$^{+}$ in g$_{9/2}$there are two seniority 4
states and one with v=2. In general though, seniority is not a good
quantum number in the g$_{9/2}$ shell (it is in the f$_{7/2}$ shell)
and so we expect the resulting eigenstates to have mixed seniority.
But Escuderos and Zamick {[}2{]} noticed that with any isospin conserving
interaction one eigenstate emerged independant of the interaction.
This was a seniority v=4 state that did not mix with the v=2 state
or indeed with the other v=4 state. There is also such a special T=2
state for J=6$^{+}$. This has lead to many theoretical works which
discuss this and prove it {[}3-8{]}. We cite in particular the nice
work by Qi {[}6{]}.

This special unique v =4 state is usually discussed in the channel
of 4 identical particles in terms of fractional parentage coefficents.
Is is displayed as such in refs {[}1{]} and {[}2{]}. But now we see
the double analog of this T=2 state in Table II in the 2proton-2 neutron
channel. We can now understand why D$^{4,2}$(0,4) is zero. If the
2 protons couple to J$_{P}$=0 they have seniority v=0. The 2 neutrons
have v=2 and so the total senioriity must be 2. However since this
state has v=4 we must have D$^{4,2}$(0,4)=0. We emphasize again that
this unique T=2 v=4 state emerges for any interaction,not just Q.Q.

But there is a new feature, as the subheading implies. Note that for
the T=0 member of the pair the amplitude D$^{4,0}$(0,4) also vanishes.
This suggests that the T=0 degenerate partner of the unique J=4 T=2
state might also have seniority v=4. In general states with both valence
neutons and protons do not have good seniority. However the J=0 T=1
pairing interaction of Flowers and Edmonds {[}9,10{]} does yield states
of good seniority. We perform a matix diagonalization for J=4$^{+}$
states in the g$_{9/2}$ shell consisting of 2 protons and 2 neutrons.
The two- body matrix elements for J=0 to J=9 are, in MeV),\{ -1,0,0,0,0,0,0,0,0,0\}.
We can identify the quantum numbers of the states by the energies.
The formula of Flowers and Edmonds {[}9,10{]} is

E=-C$_{0}$\{ (n-v)/2{*}(4j+8-n-v) -T(T+1)+t(t+1)\} 

where n is the number of nucleons (4 in this case), T is the total
isospin (zero in this case) , t is the reduced isospin, and v is the
seniority. The scale facto C$_{0}$= - 1 /(2j+1) (-1/10 in this case).
The excitation enegy relative to the J=0 ,T=0,t=0 ground state is
E{*}= E+2.2 MeV. This method of otaining quantum numbers from energies
was previously used by Neergaard {[}11{]} and by Harper and Zamick
{[}12{]}.

In this model space there are 16 J=4$^{+}$ states: 7 T=0, 6 T=1 and
3 T=2 states. We can identify the isospins in several ways, one of
which is to change the two -particle matrix element set to \{-1,-5,0,-5,0,-5,0,-5,0,-5\}
That is, we add -5 MeV to the odd J (T=0) 2-body matrix elements.
This will not affect the wave functions but will shift the states
upward by an amount 5(T{*}(T+1). The unshifted states have T=0. This
is of course also the method to separate the originally degenerate
T=0, T=2 pairs. 

Limiting ourselves to J=4 T=0 states we find the lowest excitation
energy is 1.0 MeV and the other 6 states are degenerate at 2.2 MeV.
From the quatium numbers (T,t,v) we find that the lowest state has
quantom numers (0,1,2) and the other six have (0,0,4). The first thing
that comes to mind is to see if the T=0 wave function in Tablw II
matches any of the v=4 states.However this will not work because theere
is a six fold degeneracy here . Any linear combination os these degenerate
states is also a v=4 state. However life is made simple by the fact
that there is only one v=2 state.We give the explicit wave function
in Table V. We can now take the overlap of the T=0 wave function in
Table II with the v=2 wave function in table V. We find this overlap
is zero. This clinches the fact that the T=0 wave function in Table
II has seniority v=4 (as does it's T=2 partner).

The Q.Q interaction in general does not yield states of good seniority
not only for mixed systems but even for identical particles . Here
however we have one rare exception--the T=0 partner of the T=2 unique
state. We do not get analogous degenerate pairs for J=6$^{+}$. We
did not find any degenerate pairs for 2 protons and 2 neutrons in
the i$_{13/2}$ shell or the j$_{15/2}$ shell.

We note that most interactions which conserve seniority for particles
of one kind do not conserve seniority for configurations containing
both valence protons and valence neutrons e.g. the delta interaction.
Whereas the delta interaction conserves seniority for say neutrons
in the g$_{9/2}$ shell and higher it will not conserve seniority
for the 2 proton-2 neutron configurations considerehere. In contrast
the J=0$^{+}$ T=1 pairing interaction of Flowers and Edmonds {[}9,10{]}
will yield states of good seniority for mixed systems .

In summary , with the Q.Q interaction for a configuration of 2 protons
and 2 neutrons we find selected degenerate pairs in which one member
has isospin T=0 and the other T=2. We have 4 such examples -one in
the f$_{7/2}$ shell, two in g$_{9/2}$ and one in h$_{11/2}$. The
most intriguing case involves a J=4$^{+}$ pair in the g$_{9/2}$
shell in which both members are pure seniority v=4 states. This was
anticipated for the T=2 member but is a surprise for the T=0 member.

The important message in this work is that degeneracies can lead to
confusions. When a T=0 and T=2 state are degenerate any linear combination
of the 2 states is also an eigenstate. The wave functions that appear
are each mictures of T=0 and T=2 and so the isospin information is
lost. Another example concerns the unique T=2 v=4 state for J=4$^{+}$
which was found by Escuderos and Zamick{[}1{]}. The reason it took
so long to find that such a state existed was that fractional parenage
coeffcients were generally obtained using the J=0 T=1 pairing interaction
of Edmund and Flowers {[}9,10{]}. With such an interaction the two
T=2 v=4 states are degenerate . Thus an arbritrary linear combination
of tho states can appear in a given calculation none of which looks
like the later to be dicovered unique state. Also one needs a seniority
violating interaction to find that there is a special state that nevertheless
maitains its pure (v=4) seniority. Yes, degenracies can be confusing
and we have here shown what steps have tobe taken to handle them.

I thank Kai Neergaard for useful conversations and supportive calculations.

\section{Appendix}

We show in Table VI the two-body matrix elements of Q.Q used in this
work. We list them successively from J=0 to J=2j.The even J matrix
elements have isospin T=1 whilst the odd ones have T=0.

Table VI Two- body matrix elements of the Q.Q interaction.

\begin{tabular}{|c|c|c|c|}
\hline 
J & f$_{7/2}$ & g$_{9/2}$ & h$_{11/2}$\tabularnewline
\hline 
\hline 
0 & -1 & -1 & -1\tabularnewline
\hline 
1 & -0.8095 & -0.8788 & -0.9161\tabularnewline
\hline 
2 & -0.4667 & -0.6515 & -0.7554\tabularnewline
\hline 
3 & -0.0476 & -0.3485 & -0,5325\tabularnewline
\hline 
4 & 0.3333 & -0.0152 & -0.2687\tabularnewline
\hline 
5 & 0.5238 & 0.2879 & 0.0070\tabularnewline
\hline 
6 & 0.3333 & 0.4848 & 0.2587\tabularnewline
\hline 
7 & -0.4667 & 0.4848 & 0.4434\tabularnewline
\hline 
8 &  & 0.1818 & 0.5150\tabularnewline
\hline 
9 &  & -0.5455 & 0.4026\tabularnewline
\hline 
10 &  &  & 0.0549\tabularnewline
\hline 
11 &  &  & -0.6044\tabularnewline
\hline 
\end{tabular}


\begin{thebibliography}{10}
\bibitem[1]{zamick}A. Escuderos and L. Zamick,Romanian Journal of
Physics, Vol. 58, Nos. 9-10, pp 1064-1075, (2013)$_{P}$

\bibitem[2]{key-1}A. Escuderos and L.Zamick, Phys. Rev. C73 , 044302
(2006)

\bibitem[3]{key-2}A. Escuderos and L.Zamick, Ann. Phys. (NY)321,987
(2006)

\bibitem[4]{key-1}L.Zamick and P. Van Isacker , Phys. Rev. C78, 044327
(2008).

\bibitem[5]{key-1}I. Talmi, Nucl.Phys.A 846(2010) 31

\bibitem[6]{key-2} Chong Qi ,Phys.Rev.C83 014307,2011

\bibitem[7]{key-3}Chong Qi, Z.X.Yu, R.J. Liotta,Nucl Phys. A 884-885
(2012) 21

\bibitem[8]{key-2}P.Van Isacker and S. Heinze, Annals of Physics
349 (2014) 73

\bibitem[9]{key-1}B.H. Flowers, Proc. Roy. Soc. (London)A212, 248
(1952)

\bibitem[10]{key-2}A.R. Edmonds and B.H. Flowers, Proc. Roy. Soc.
(London) A214,515 (1952)

\bibitem[11]{key-1}K.Neergaard, Phys. Rev. C90,014318 (2014)

\bibitem[12]{key-2}M. Harper and L.Zamick Phys. Rev.C91, 014304 (2015)\end{thebibliography}
\end{document}